\documentclass[conference,compsoc]{IEEEtran}
\ifCLASSOPTIONcompsoc
  \usepackage[nocompress]{cite}
\else
  \usepackage{cite}
\fi
\ifCLASSINFOpdf
\else
\fi
\usepackage{url}

\usepackage{xcolor}
\usepackage{listings}
\usepackage{graphicx}

\begin{document}
%
\title{comokit4py : a python package to ease COMOKIT agent based model simulation integration into a high performance computing workflow}

\author{
\IEEEauthorblockN{Arthur Brugière}
\IEEEauthorblockA{UMMISCO UMI 209, SU/IRD \\
Bondy, France \\
Thuyloi University \\
175 Tay Son, Dong Da, Hanoi, Vietnam \\
Email: arthur.brugiere@ird.fr}
\and
\IEEEauthorblockN{Kevin Chapuis}
\IEEEauthorblockA{ESPACE-DEV, Univ Montpellier, IRD, \\
Univ Antilles, Univ Guyane, Univ Réunion\\
Montpellier, France\\
Email: kevinchapuis@gmail.com}
}


%


\maketitle

\begin{abstract}
Agent-based model (ABM) are a kind of computer model that makes it possible to simulate a set of autonomous interacting programs called agents in a shared virtual environment. Among other application field, it has been commonly used to simulate social phenomena such as urban segregation, opinion dynamic or epidemiological crisis \cite{gilbert2004agent}. Recently, a research emphasis has been put on ABM to study \textit{in silico} the impact of non-pharmaceutical interventions to mitigate the SARS-CoV-2 outbreak of 2020, with few of them that had a great impact on global political responses \cite{ferguson_report_2020}. Among the model used COMOKIT \cite{gaudou_comokit_2020} has been design to simulate the every-day-life of inhabitant of various cities in Vietnam and test policy interventions for various COVID-19 spread scenarios. Such endeavor required huge computational power to handle a huge number of simulation replication over a large set of parameters. In this proposal we present a python package that enables to easily generate, explore and build reports for any COMOKIT experiment to be launched over High-Performance Computing (HPC) infrastructure.
\end{abstract}


%
\IEEEpeerreviewmaketitle

\section{Introduction}

Agent-based models are a class of computational models of Artificial Intelligence that consist in bringing together active and autonomous computing entities interacting in a shared virtual environment. These entities called agents, which are most often represented as an autonomous program composed of algorithms (i.e., their behaviors) and variables (i.e., their personal characteristics), are characterized by a set of internal states that determine their reasoning and behavior. The interaction between agents, made possible by the definition of a shared global environment, is at the heart of the agent-based model dynamics: the ability for communication, cooperation, and modification of the environment are among the actions available to the agents to achieve their own or collective objectives. It is through their capabilities for action and interaction with other agents and the environment, that these entities determine with more or less autonomy — a spectrum ranging from a purely reactive agent to a socio-cognitive agent — their internal states and their reasoning.

Among applications of such kind of models, social simulation is of high importance in the field
. In those ABM, the geographical environment of the agent is a structuring element of interactions and decisions taken individually and collectively: it can represent a city, with its road or communication networks, or a natural environment, with characteristic spaces (e.g., forest, meadow, mountain) and dynamics (e.g., hydrological, atmospheric) 
. The parallel between agent-based models and human society makes it possible to represent in a symbolic way the actors, their common environment and their interactions for a potentially unlimited number of social issues. 

The major advantage of this type of approach is that it allows modelers to experiment, through simulation, with a set of scenarios that are impossible or too costly to test or reproduce in reality: mass evacuation exercises \cite{daude_escape:_2019}, large-scale urban development \cite{brown2006effects}, corporate reorganization \cite{chapuis_happywork_2016}, or the introduction of intervention policies during a major health crisis \cite{gaudou_comokit_2020}.

However, to be as descriptive as possible, these models often includes many processes that requires a lot of data and a huge number of parameters. Furthermore, when data is missing and the processes modeled lack of knowledge foundation, they often rely on stochastic mechanisms. These two points are a problem when it comes to support decision-making based on outcomes of such models. 

The solution to the latter, i.e. stochastic nature of simulations, mainly consists in running many replications of the model and compiles the outcomes to reduce the impact of randomness. The first problem, also known as the \textit{curse of dimensionality}, namely the practically impossible exhaustive exploration of every combination of parameters' values, is decomposed in several type of sensitivity analysis, e.g., local, global, with or without exhaustiveness, with or without optimization. In both cases, the process that consists in generating, gathering and thoroughly analyzing results from a model is truly process consuming.

A simple laptop isn't enough to handle such demanding task, this is why the exploration of descriptive and social agent-based models is usually done on \textit{High Performance Computing} (\textit{HPC}). Those are servers dedicated and specialized in computationally intensive tasks and used in many fields. This article presents a python package that makes it easy to explore an agent-based model on top of an HPC workflow.

The article is structured as follows: Section \ref{sec:comokit} introduce the agent-based model called COMOKIT for which this package has been made, Section \ref{sec:abms_x_hpc} exposes why and how HPC is used to explore agent-based simulation models, Section \ref{sec:comokit4py} details the content and usage of the proposed package while Section \ref{sec:comokit_analysis} demonstrates its application to a sensitivity analyzes problem; Section \ref{sec:conclusion} concludes our proposal and open toward future developments.

\section{The COMOKIT model}\label{sec:comokit}

COMOKIT is an agent-based model of COVID-19 spread at the scale of a city. It has been implemented by an interdisciplinary group of researchers using the GAMA-Platform \cite{taillandier2019building} and the GAML (\textit{GAMa Language}) programming language. The model features three main sub-models of (i) individual clinical state dynamics and agent-to-agent epidemiological transmission based on a classical SEIR model, (ii) realistic synthetic population with daily mobility at a one-hour scale and (iii) lastly, policy intervention to mitigate propagation. Each module is highly flexible and can be extended in several ways. A complete description (following the ODD protocol) is available in \cite{gaudou_comokit_2020}.

The model has been developed as part of the Vietnamese Quick Response Team to support intervention in the field. It has been used to model and extrapolate the impact of mitigation policy such as test \& trace, quarantine and lockdown, in several cities in Vietnam (e.g., Hanoi, Son Loi, Ben Tre).

To explore the potential impact of interventions to mitigate the propagation, the COMOKIT team relied on HPC to conduct large experiments. Among the problems that arise, parameters calibration, exploration, and simulation replications required days of computation to be able to align simulation outcomes with decision-making issues. The pyhton packages describe in the article has been part of the work effort carried out by members of that team.

 

\section{ABMS platforms deployment on HPC}\label{sec:abms_x_hpc}


Contrary to Deep learning, \textit{ABM Simulation} (\textit{ABMS}) do not look after an optimization of the simulation execution but rather paralleled replications and simulations to speed up the whole process of parameter space exploration. In order to achieve this parallelization, three different categories of software solutions exist:

\subsection{With default ABMS software}

The first solution is the direct use of ABMS \textit{Integrated Development Environment} (\textit{IDE}) software programs with a server mode (commonly called \textit{headless mode}). This IDE mode provide the ability to compute models on the server side.


The \textit{GAMA Platform} \cite{taillandier2019building} or the \textit{NetLogo} software \cite{tisue2004netlogo} are ABMS IDE which provide a headless mode to allow heavier model's exploration on server-side execution. However, their usage may be limited and can require tools to be correctly used in an HPC environment (cf \ref{ssec:wrapper}).

Some other software programs enhance their headless mode to fully integrate the HPC's execution eco-system (with the support of job scheduler like SLURM \cite{yoo2003slurm} or native support of large-scale distributed computing systems execution) in their headless mode. That's what proposes \textit{Repast HPC} \cite{repast_hpc_2013} for instance.
\vspace{-.1cm}
\subsection{With dedicated software}
\vspace{-.1cm}
The second category requires the use of external (from the default IDE, or any kind of simulation engine) software using any simulation engine to drive a given exploration plan of the model. This monitoring can be simple (exhaustive exploration) or using different algorithms to optimize the explored parameter space and, so, the number of simulation. The second advantage of this solution is that, these software programs are specially developed to run on HPC and fully optimize the exploration on the decentralized architecture of these servers.


The \textit{OpenMole}\cite{REUILLON20131981} platform or \textit{EMEW} \cite{ozik2018high}, for instance, allow to easily use complex exploration algorithms and apply them on model developed in several languages.

While this may seems like a convenient solution, such software often uses custom languages that must be mastered before the full potential of this solution can be realized.
\vspace{-.1cm}
\subsection{With software wrapper}\label{ssec:wrapper}
\vspace{-.1cm}
The last category is what is commonly known as \textit{wrapper libraries}. These are small packages in common programming languages (R, Python, etc.) that allow you to easily manage another program within these common languages. Applied in our case, these libraries allow to manage the headless part of an ABMS software (first category) to launch exploration from a custom script without having to know nor understand this IDE usage.

Among all the existing ones, we can cite \textit{Nl4py} \cite{gunaratne2018nl4py} allowing to pilot the IDE \textit{NetLogo} in Python's scripts, and \textit{GAMAR} \footnote{Project in development: https://github.com/r-and-gama/gamar/} allowing to use R to pilot the \textit{GAMA Platform}.

This last method seemed, in the COMOKIT project, the most convenient way to explore models. It can be adapted to fit heterogeneous server architectures used within the project and split the development keeping the heterogeneous scientific-development team in each other expertise, accelerating, overall, the development of the project.


\section{comokit4py: deploy COMOKIT on HPC}\label{sec:comokit4py}

comokit4py is a software wrapper developed in the COMOKIT project, wrapping the GAMA Platform for exploring agent-based models in an HPC architecture.

The comokit4py library usage relies on three main classes : \texttt{Gama} to set the GAMA-platform on the server, \texttt{GamaExploration} to define the exploration plan and \texttt{Workspace} to launch the headless exploration on the HPC.

\subsection{Set the GAMA-Platform}\label{ssec:setupGama}

The first object to define is \texttt{Gama} which will define which installed version of the GAMA-Platform to wrap with this script and use in the exploration.
The few parameters to configure are the path (relative to the script or absolute path, the library will automatically convert it in absolute path) to the \texttt{gama-headless.sh} default headless launch script (\texttt{pathToHeadlessScript}) and the memory to allow at the GAMA headless' JVM (\texttt{memory}).

An implemented example is :

\begin{lstlisting}[language=Python]
import os, comokit4py

gamaPathHeadless = "./GAMA_1.8.1_Linux/ headless/gama-headless.sh")

gama = comokit4py.Gama( pathToHeadlessScript = gamaPathHeadless, 
    memory = "8go")
\end{lstlisting}

\subsection{Define the exploration plan}\label{ssec:definePlan}


The second object to define is the definition of the exploration plan with the \texttt{GamaExploration} class.

In this object, user should define the name of the COMOKIT-GAML's experiment to explore (\texttt{experimentName}), the path to the \textit{gaml} file in which is coded the experiment (\texttt{gamlFile}), the number of replication of the simulation in the exploration (\texttt{replication}), the limit number of step in the simulation before forcing to stop it (\texttt{final}).

Some other optional parameter can be used, e.g. the number of experiment per XML file (the interest of it will be explained later — \texttt{experimentPerXML}), a boolean ending condition (to prevent to reach the step limit condition), and a starting seed value (to expand a previous experiment exploration).

At the end, the object's function \texttt{calculatesExperimentSpace} should be triggered to parse the gaml file and automatically generate the full experiment space to explore (i.e., every parameter combination).

An implemented example is :

\begin{lstlisting}[language=Python]
import comokit4py

explo = comokit4py.GamaExploration( experimentName = "Headless",
    gamlFile = "./COMOKIT/Model/COMOKIT /Experiments/Physical Interventions/Significance of Wearing Masks.gaml",
    replication = 1000,
    final = 5000,
    experimentPerXML = 8)

explo.calculatesExperimentSpace()
\end{lstlisting}

\subsection{Set up the exploration workspace}\label{ssec:explorationWS}

The final object, \texttt{Workspace}, manage the exploration execution with the 2 previously defined objects and some extra variables as the path to the headless logs (\texttt{workspaceDirectory}). An optional parameter allows rebuilding, in the HPC file system, all the work space architecture used by comokit4py to clean it.

Once the object defined, as for the \texttt{GamaExploration} object, an object's function should be called to generate the needed files for the exploration. In this case, it'll create all the precalculated experiment space (cf. \ref{ssec:definePlan}) in XML files which will be used by the GAMA headless. These XML files are central to the use of headless because they represent the input files of GAMA and include the definition of a batch of simulations to be executed.

An example usage, following the two-previous definition (\ref{ssec:setupGama} and \ref{ssec:definePlan}), is :

\begin{lstlisting}[language=Python]
#   [...]

ws = comokit4py.Workspace(
    gama = gama, 
    explorationPlan = explo, 
    workspaceDirectory = "./out")

ws.generateNeededForExploration()
\end{lstlisting}

\subsection{Running headless exploration}

The library allows to launch the exploration in two different modes: the first one is with the job scheduler \textit{SLURM} \cite{yoo2003slurm}, and the second is a basic GAMA headless wrapper  mode.


EDF support the COMOKIT project\footnote{\url{https://www.edf.fr/groupe-edf/qui-sommes-nous/activites/recherche-et-developpement/toutes-les-actualites-de-la-r-d/la-r-d-d-edf-et-des-equipes-de-l-ird-s-allient-dans-la-lutte-contre-le-covid-19}} by allowing the executing on one of their HPC called GAIA\footnote{\url{https://www.top500.org/system/179569/}}. This supercomputer use the job scheduler called \textit{SLURM} which dispatch every simulation on the hardware infrastructure.

comokit4py has been firstly developed to enhance and facilitate this usage (with GAIA) and provide functions which generate extra files for this use case.
First, it will be important to enable the private variable \texttt{edf} which will set some meta-data for using GAIA with the function \texttt{Workspace.setEdfBool()}.
Thereafter, the function \texttt{Workspace.prepareSBatch()} will be called to generate needed resources allowing the proper usage of SLURM in batch mode. These files can be set with a timeout value for each job (\texttt{jobTimeout}), some parameters for the HPC usage (the number of core per node (\texttt{core}), number of nodes used (\texttt{nodes}), the max number of concurrent submissions (\texttt{maxSubmission})).
Finally, the function \texttt{Workspace.runSlurm()} will use these files to start a SLURM submission and launch the exploration.
Below, an example with the configuration used with GAIA and following the previous code sample (\ref{ssec:explorationWS}):

\begin{lstlisting}[language=Python]
[...]
ws.generateNeededForExploration()

ws.setEdfBool(True)
ws.prepareSBatch(jobTimeout = 7, 
    core = 36, nodes = 16, 
    submission = 1, maxSubmission = 6)

ws.runSlurm()
\end{lstlisting}

In a simpler execution mode, it is also possible to launch an exploration without using the SLURM scheduler. In that case, the python script will run every XML files by starting one instance of GAMA at a time and parallelizing one simulation per core allocated.

The code to explore with that mode is, after generating the GAMA's exploration files (\ref{ssec:explorationWS}), launching the software with the number of cores allocated to the JVM (by default, it's using all cores available).
Below, a usage example:

\begin{lstlisting}[language=Python]
[...]
ws.generateNeededForExploration()

ws.runGamaHeadless(cores = 4)
\end{lstlisting}

\subsection{Processing output data}

Once the exploration done, the \textit{batch\_output} folder will be filled up with a significant amount of raw data stored in CSV files. This last step aim at turning Gigabytes of data into few MiB humanly readable data.

This is achieved by aggregating replications to obtain a synthetic output to assess what happens at a certain point in the parameter space. Secondly, backed by the reduced amount of data generated by simulation replications, data are processed and converted (in CSV or PNG output file format, cf figures \ref{ssec:result}) to have a direct view of the results, verification, and insights to better understand the outcome of the model.

Within the comokit4py library, the first step concatenate and processed the raw data with several variables, and they can be exported in various file format as showed in the example below:

\begin{lstlisting}[language=Python]
[...]
ws.prepareProcessedOutput(quartile = True, startDate = [2020, 1, 24], startPolicyDate = [2020, 3, 17], endPolicyDate = [2020, 5, 11])

p = ws.rawOutputProcessing()

ws.generateCsv(output = p)
ws.generatePng(title = "Title", output = p)
\end{lstlisting}

\section{comokit4py: an application to sensitivity analysis}\label{sec:comokit_analysis}

At first, COMOKIT have been applied to explain and foreseen potential consequences of various political interventions in the Son Loi commune
. Global initialization over GIS and demographic data, as well as parameter values, and intervention policies tested and explored have been thoroughly discussed in \cite{gaudou_comokit_2020}. However, except for the stochastic impact over simulation outcome, sensitivity analysis has been replaced by parameter default values taken from the literature. The main reason was to provide a quick response to support decision-making in the crisis context of the very first outbreak peak of March 2020.

However, in the meanwhile we have been trying to explore the impact of parameters that have been arbitrary set to default value based on available knowledge we got at that time. As it is the case for most ABM, exploring the impact of all the combination of parameter values is practically impossible \cite{lee2015complexities,ten2016sensitivity}. For example, the epidemiological module of COMOKIT has more than 58 parameters with most of them being continuous variables; let's assume that we account for 10 values each, it means 58$^{10}$ point in the parameter space, hence leading to 4.3e+17 simulation runs multiplied by the number of replications.

\subsection{Explore the impact of a potential environmental contamination in the spread of SARS-CoV-2}\label{ssec:result}

In the context of this paper we propose to exemplify the use of comokit4py on a subset of parameter of interest that have been ignored in other COMOKIT experiments, but still of high importance to understand and explore solutions to overcome the current crisis: this mechanism is the potential environmental contamination of human due to persistence of the SARS-CoV-2 on physical surfaces \cite{chin2020stability}. In COMOKIT the spatial unit of agent infection, i.e., building, can be contaminated by infectious agent, and then infect back other susceptible agents that are or will be present.

To explore the impact of a potential environmental contamination, we identified two key parameters: a continuous value that represent the amount of viral load an individual can release in the environment (\texttt{basic\_viral\_release} $\in [0.01;0.1]$) and decrease factor of this viral load contained in the environment (\texttt{basic\_viral\_decrease} $\in [0.02;0.2]$). Successful transmission from building to individual is based upon a probability linearly deduced from the aggregation of all individual viral load released on the environment. The decreased of the accumulated viral load in buildings are ranged from 2\% to 20\% per hour.

\subsection{Build experiment to execute on HPC, extract and synthesis data to visualize results}

The experiment is set to explore a finite number of values for the two parameters, with each numerical value in a range from min to max value included, with a step equal to $1/10$ of the max value. This makes an experiment plan that should include 10 values for parameter \texttt{basic\_viral\_release} and 10 values for \texttt{basic\_viral\_decrease}, so a total of 100 points in the parameter space. For testing the limit of the library, this plan will launch 1000 replications for each parameters' combination, which makes an overall charge of 100k simulations to dispatch over 89 nodes on Gaia HPC. This took around 31 hours and produced more than a Terabyte of raw data.

\begin{figure}[hbtp]
    \centering
    \includegraphics[width=\columnwidth]{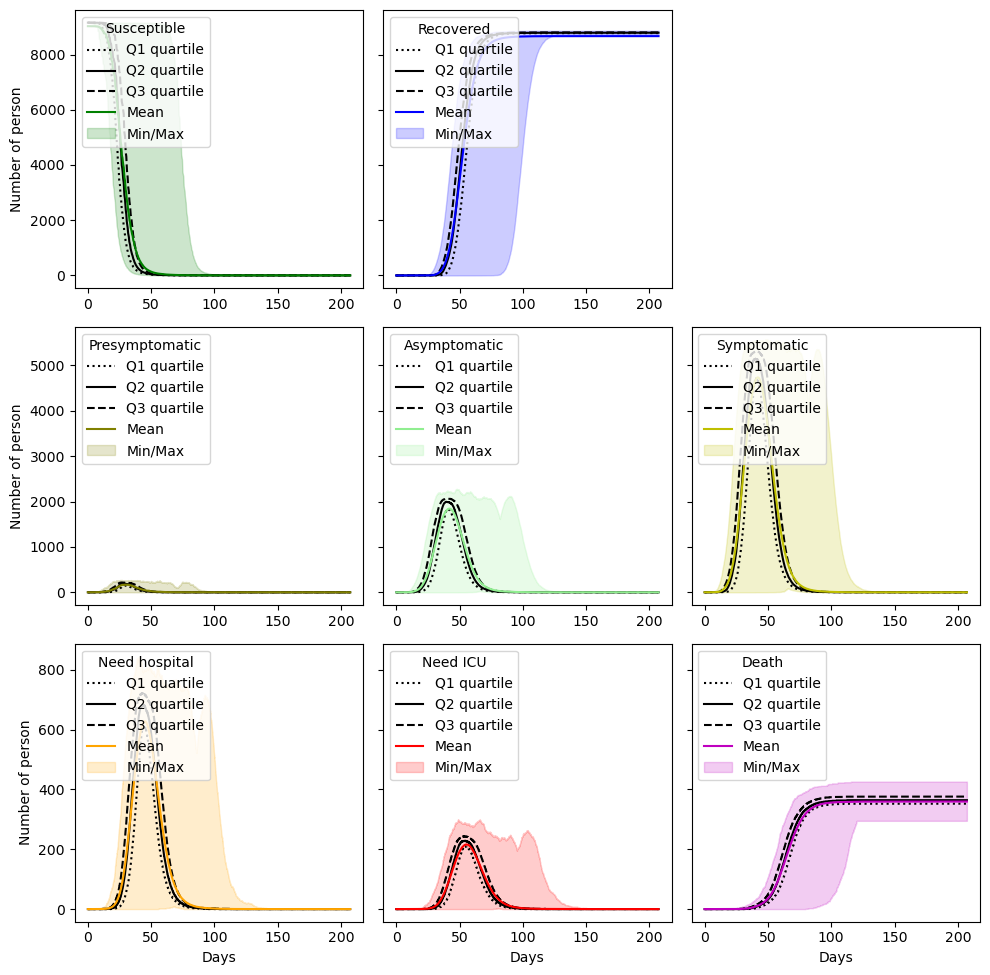}
    \caption{Overall view of a simulation with replications aggregation and over the indicators of interest}
    \label{fig:graphs}
\end{figure}

First aggregation and compilation of raw data leads to the construction of simulation results for each COMOKIT's simulation indicators. The graph \ref{fig:graphs} exhibit the visual output of this first stage.

The result displays the quartiles (lines for Q1, Q2 and Q3), min and max values (colored interval) over simulation replications in respect to time step (x-axis), for a total of 8 indicators of interest: first row of \ref{fig:graphs} depicts values relative to susceptible and recovered, second row shows the values for presymptomatic, asymptomatic and symptomatic, while last row displays values for the number of individual in hospital, number in intensive care unit (\textit{ICU}) and cumulative number of deaths.

These set of graphics makes it possible to have simulation trends of one point of the parameter space over time, for any number of replications. Hence, we can go into details of each explored point in the parameter space and investigates different simulation dynamics considering specific conjunction of parameter values. However, as it goes into the details of each point of the exploration, it fails to show in a glance 
how parameters interactions may change the dynamic of the simulated model.

\begin{figure}
    \centering
    \includegraphics[width=\columnwidth]{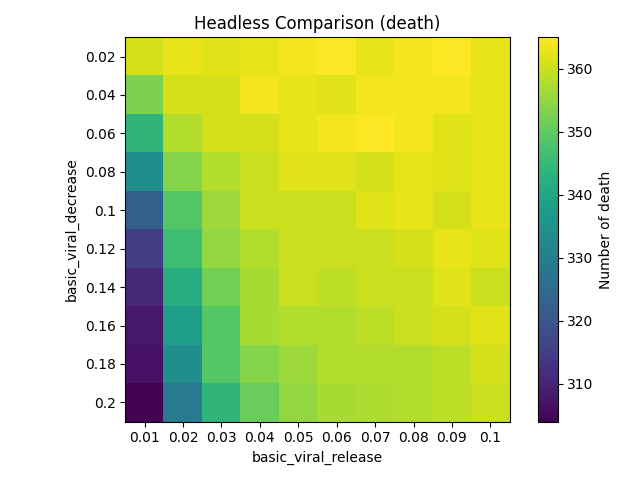}
    \caption{Median number of death over 1000 simulations replications over every values of individual viral release (x-axis) and building viral load decrease (y-axis)}
    \label{fig:number_death}
\end{figure}

\begin{figure}
    \centering
    \includegraphics[width=\columnwidth]{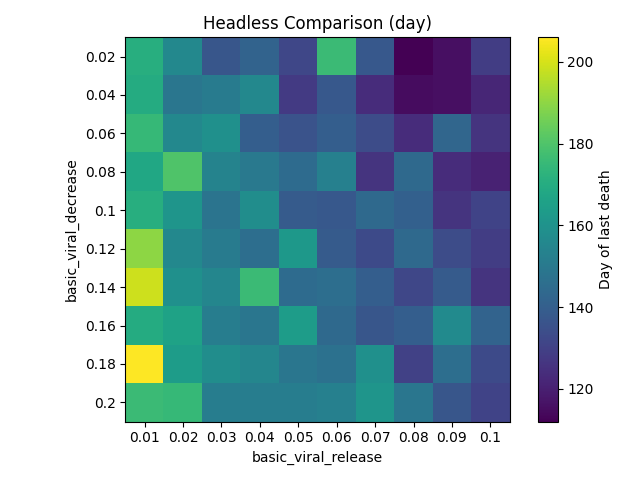}
    \caption{Median number of last day with death over 1000 simulations replications with different value of individual viral release (x-axis) and building viral load decrease (y-axis)}
    \label{fig:day_death}
\end{figure}

In fact, considering parameter exploration this is of crucial interest to have a synthetic view over the explored space, to elicit trends and global dynamics when several parameter value changes \cite{ten2016sensitivity}. In the two graphics \ref{fig:number_death} and \ref{fig:day_death} generated by comokit4py, we have been able to gauge the outcome of simulations on one indicator of interest for every explored point of the parameter space.

The figures \ref{fig:number_death} and \ref{fig:day_death} represents the cumulative number and last day of death respectively, considering each possible value for the individual viral release in the environment (x-axis = \texttt{basic\_viral\_release}) and the decrease speed of viral load contained in buildings (y-axis = \texttt{basic\_viral\_decrease}).

We can clearly outline the impact of each parameter on the model outcomes. For instance, with low individual viral release values, the environmental decrease of SARS-CoV-2 presence on surfaces has a great impact, leading to potentially increase the number of death from 304 to 361 (16\% increase); however, for reasonably high value of individual release impact of decrease value remain marginal ($<$ 2.5\%). The influence of parameter changes on the outbreak timeline is even stronger: conjunction of the highest individual viral release (e.g., without mask) and low environmental decrease will lead to an outbreak that sustains 45.1 days longer on average than for low to very low environmental contagion\footnote{we compare combined results of 3 highest with lowest values for both parameter}.



\section{Conclusion}\label{sec:conclusion}


In this paper we presented \texttt{comokit4py} a python package to ease the integration of the COMOKIT agent-based model in an HPC workflow. The proposed solution has been used to explore through the model the impact of the potential environmental contamination to SARS-CoV-2. The proposed tool makes it possible to easily deploy and analyze a large-scale agent-based simulation experiment (a total of 100k simulations) and have proven to be of great interest to understand one of the key features of the COMOKIT model.

As a mention, our solution remains perfectible. We plan to extend the content of the package toward more flexibility in aggregation and visualization features, making it possible to specify aggregation formula as well as post-computed outcome of interest. Another big way to extend this tool is to work toward a generalization of the process to handle more models developed on the GAMA platform.  


\ifCLASSOPTIONcompsoc
  \section*{Acknowledgments}
\else
  \section*{Acknowledgment}
\fi

This is part of the COMOKIT project funded by french ANRS agency (project COV15, AAP COVID19-Sud Flash 2020) and a research collaboration between IRD and EDF R\&D



\bibliographystyle{IEEEtran}
\bibliography{main.bib}

\end{document}